 \documentclass[12pt,preprint]{aastex}

 \usepackage{graphics}
\usepackage{graphicx}
\slugcomment{Accepted for publication in Astrophysics \& Space Science}

\shorttitle{Trajectory and stability regions\dots Chermnikh-Like problem}
\shortauthors{Badam Singh Kushvah}

\begin{document}


\title{Trajectory and stability  of Lagrangian point $L_2$ in the Sun-Earth system}


\author{Badam Singh Kushvah}
\affil{Department of Applied Mathematics, Indian School of Mines Dhanbad-826004(Jharkhand) India}
\email{bskush@gmail.com,kushvah.bs.am@ismdhanbad.ac.in}

%\altaffiltext{1}{SERC-DST New Delhi (India) Fast Track  Young Scientist}
%CTIO is operated by AURA, Inc.\ under contract to the National Science
%\altaffiltext{2}{Society of Fellows, Harvard University.}
%\altaffiltext{3}{present address: Center for Astrophysics,
%    60 Garden Street, Cambridge, MA 02138}
%\altaffiltext{4}{Visiting Programmer, Space Telescope Science Institute}
%\altaffiltext{5}{Patron, Alonso's Bar and Grill}


\begin{abstract}
This paper describes  design of the trajectory and analysis of  the stability of collinear point $L_2$  in the  Sun-Earth system.  The modified restricted three body problem with additional gravitational potential from the  belt  is used as the model for the Sun-Earth system. The effect of   radiation pressure of the Sun and oblate shape of the Earth are considered.  The  point $L_2$ is asymptotically stable  upto  a specific value  of time $t$  correspond to each set of  values of  parameters and initial conditions.   The results  obtained from   this study would be applicable  to locate a satellite, a telescope  or a space station around the point $L_2$.
\end{abstract}





\section{Introduction}
\label{Intro} 
This paper deals with the Sun-Earth system with the modified  restricted
three body problem model[as in \cite{Kushvah2009Ap&SS,Kushvah2009RAA}] including   radiation pressure,
oblateness of the Earth and influence of the belt. Further it considered that
the  primary bodies are moving in circular orbits about their center
of mass. It is  well-known that  five equilibrium points(Lagrangian points)
that appear in the planar restricted three-body problem are very
important for astronautical applications. The collinear points are
unstable and the triangular points are conditionally stable in the classical restricted three body problem[please see \citet{Szebehely1967}].
This can be seen in  the Sun-Jupiter system  where several thousand asteroids(collectively
referred to as Trojan asteroids), are in orbits of triangular
equilibrium points. But collinear equilibrium points are also made
linearly stable by continuous corrections of their orbits(\lq\lq
halo orbits\rq\rq). In other words the collinear equilibrium points
are metastable points in the sense that, like a ball sitting on top
of a hill. However, in practice these Lagrange points have proven to
be very useful indeed since a spacecraft can be made to execute a
small orbit about one of these Lagrange points with a very small
expenditure of energy[please see \citet{Farquhar1967JSpRo,Farquhar1969AsAer}].

We considered the Chermnykh's problem  which is a new kind of restricted three body
problem, it   was first time studied by \citet{Chermnykh1987}. This
problem generalizes two classical problems of Celestial mechanics:
the two fixed center problem and the  restricted three body problem.
This gives wide perspectives for applications of the problem  in
celestial mechanics and astronomy. The importance of the problem in
astronomy has been addressed  by \citet{Jiang2004IJBC}. Some
planetary systems are claimed to have discs of dust and they are
regarded to be young analogues of the Kuiper Belt in our Solar
System. If these discs are massive enough, they should play
important roles in the origin of planets\rq orbital elements. Since
the belt of planetesimal often exists within a planetary system and
provides the possible mechanism of orbital circularization, it is
important to understand the solutions of dynamical systems with the
planet-belt interaction.The Chermnykh's problem  has been  studied by
many scientists such as \citet{JiangYeh2004AJ},
\citet{Papadakis2004A&A},\citet{Papadakis2005Ap&SS299} and \citet{JiangYeh2006Ap&SSI,YehJiang2006Ap&SSII}.

The present paper investigates the nature of collinear
equilibrium point $L_2$ because of  the interested point  to  locate an artificial satellite. Although there are two new equilibrium points due to
mass of the belt(larger than 0.15) as obtained by \citet{JiangYeh2006Ap&SSI,YehJiang2006Ap&SSII}, but they are left to be
examined. All the results are computed numerically using same technique  as  in  \cite{Grebennikov2007CMMPh}, because pure analytical methods are not suitable.  For specific
time intervals, and initial values,  these results provide new
information on the behavior of trajectories around the  Lagrangian point $L_2$.


\section{Location of the Lagrangian Points}
 \label{EqJaco}
It is  supposed that the motion of an infinitesimal mass particle is
influenced by the gravitational force from primaries and a belt of
mass   $M_b$.  The  units of the mass and  the distance  are
taken such that sum of the masses and the distance between primaries
are unities. The unit of the time i.e. the time period of $m_1$
about $m_2$ consists of $2\pi$ units such that  the Gaussian
constant of gravitational $\mathbf{ k}^{2}=1$. Then perturbed mean
motion $n$ of the primaries is given by
$n^{2}=1+\frac{3A_{2}}{2}+\frac{2M_b
r_c}{\left(r_c^2+T^2\right)^{3/2}}$, where
$T=\mathbf{a}+\mathbf{b}$, $\mathbf{a,b}$ are flatness and  core
parameters respectively which determine the density profile of the
belt.    $r_c^2=(1-\mu)q_1^{2/3}+\mu^2$,
$A_{2}=\frac{r^{2}_{e}-r^{2}_{p}}{5r^{2}}$ is the oblateness
coefficient of $m_{2}$; $r_{e}$, $r_{p}$  are the equatorial and
polar radii of $m_{2}$ respectively.   $r=\sqrt{x^2+y^2}$ is  the
distance between primaries and $x=f_1(t), y=f_2(t)$ are the
functions of  time $t$ i.e. $t$ is only independent variable. The
mass parameter is $\mu=\frac{m_{2}}{m_{1}+m_{2}}$($9.537 \times
10^{-4}$ for the Sun-Jupiter and $3.00348 \times 10^{-6}$ for the
Sun-Earth mass distributions respectively ).  $q_1=1-\frac{F_p}{F_g}$
is a mass reduction factor, where   $F_{p}$ is the solar radiation
pressure force which is exactly apposite to the gravitational
attraction force $F_g$.  The coordinates of $m_1$, $m_2$ are
$(-\mu,0)$, $(1-\mu,0)$ respectively. In the above mentioned
reference system and \citet*{MiyamotoNagai1975PASJ} model, the
equations of motion of the infinitesimal mass particle in the $x
y$-plane formulated as[please see
\citet{Kushvah2008Ap&SS318},\cite{Kushvah2009Ap&SS}]:
\begin{eqnarray}
\ddot{x}-2n\dot{y}&=&\Omega_x ,\label{eq:Omegax}\\
\ddot{y}+2n\dot{x}&=&\Omega_y,\label{eq:Omegay}
 \end{eqnarray}
where
\begin{eqnarray*}
\Omega_x&=& n^{2}x-\frac{(1-\mu)q_1(x+\mu)}{r^3_1}-\frac{\mu(x+\mu-1)}{r^3_2}-\frac{3}{2}\frac{\mu{A_2}(x+\mu-1)}{r^5_2}\nonumber\\
&&-\frac{M_b x}{\left(r^2+T^2\right)^{3/2}},\label{eq:Omegax1}\\
\Omega_y&=&n^{2}y
-\frac{(1-\mu)q_{1}{y}}{r^3_1}
-\frac{\mu{y}}{r^3_2}-\frac{3}{2}\frac{\mu{A_2}y}{r^5_2}-\frac{M_b y}{\left(r^2+T^2\right)^{3/2}},\label{eq:Omegay1}\end{eqnarray*}
\begin{eqnarray}
\Omega&=&\frac{n^2(x^2+y^2)}{2}+\frac{(1-\mu)q_1}{r_1}+\frac{\mu}{r_2}+\frac{\mu
 A_2}{2r_2^3}+\frac{M_b}{\left(r^2+T^2\right)^{1/2}},\label{Omega}\\
r_1&=&\sqrt{(x+\mu)^2+y^2}, r_2=\sqrt{(x+\mu-1)^2+y^2}.\nonumber
\end{eqnarray}
From  equations (\ref{eq:Omegax}) and (\ref{eq:Omegay}),  the
Jacobian integral is given by:
\begin{eqnarray}
    E=\frac{1}{2}\left(\dot{x^2}+\dot{y}^2\right)-\Omega(x,y,\dot{x},\dot{y})=(\mbox{Constant}),
\end{eqnarray}
which is related to the Jacobian constant  $C=-2E$. The location of
three collinear equilibrium points and two triangular equilibrium
points  is computed by dividing the orbital plane into three parts
$L_1,L_{4(5)}$: $\mu<x<(1-\mu)$, $L_2$: $(1-\mu)<x$ and $L_3$:
$x<-\mu$. For the collinear points,  an algebraic equation of the
fifth degree is solved numerically with initial  approximations to
the Taylor-series as:
\begin{eqnarray}
x(L_1)&=&1-(\frac{\mu}{3})^{1/3}+ \frac{1}{3}(\frac{\mu}{3})^{2/3}-\frac{26\mu}{27}+\dots,\label{eq:xL1}\\
x(L_2)&=&1+(\frac{\mu}{3})^{1/3}+ \frac{1}{3}(\frac{\mu}{3})^{2/3}-\frac{28\mu}{27}+\dots,\label{eq:xL2}\\
x(L_3)&=& -1-\frac{5\mu}{12}+ \frac{1127\mu^3}{20736}+\frac{7889\mu^4}{248832}+\dots.\label{eq:xL3}
\end{eqnarray}
The solution of differential  equations (\ref{eq:Omegax}) and
(\ref{eq:Omegay}) is  presented as interpolation function which is
plotted  for various integration intervals by substituting specific
values of  the time $t$ and initial conditions i.e.
$x(0)=x(L_i),y(0)=0$ where $i=1-3$ and $x(0)= \frac{1}{2}-\mu,
y(0)=\pm\frac{\sqrt{3}}{2}$ (for the triangular equilibrium points).
\begin{figure}
   \plottwo{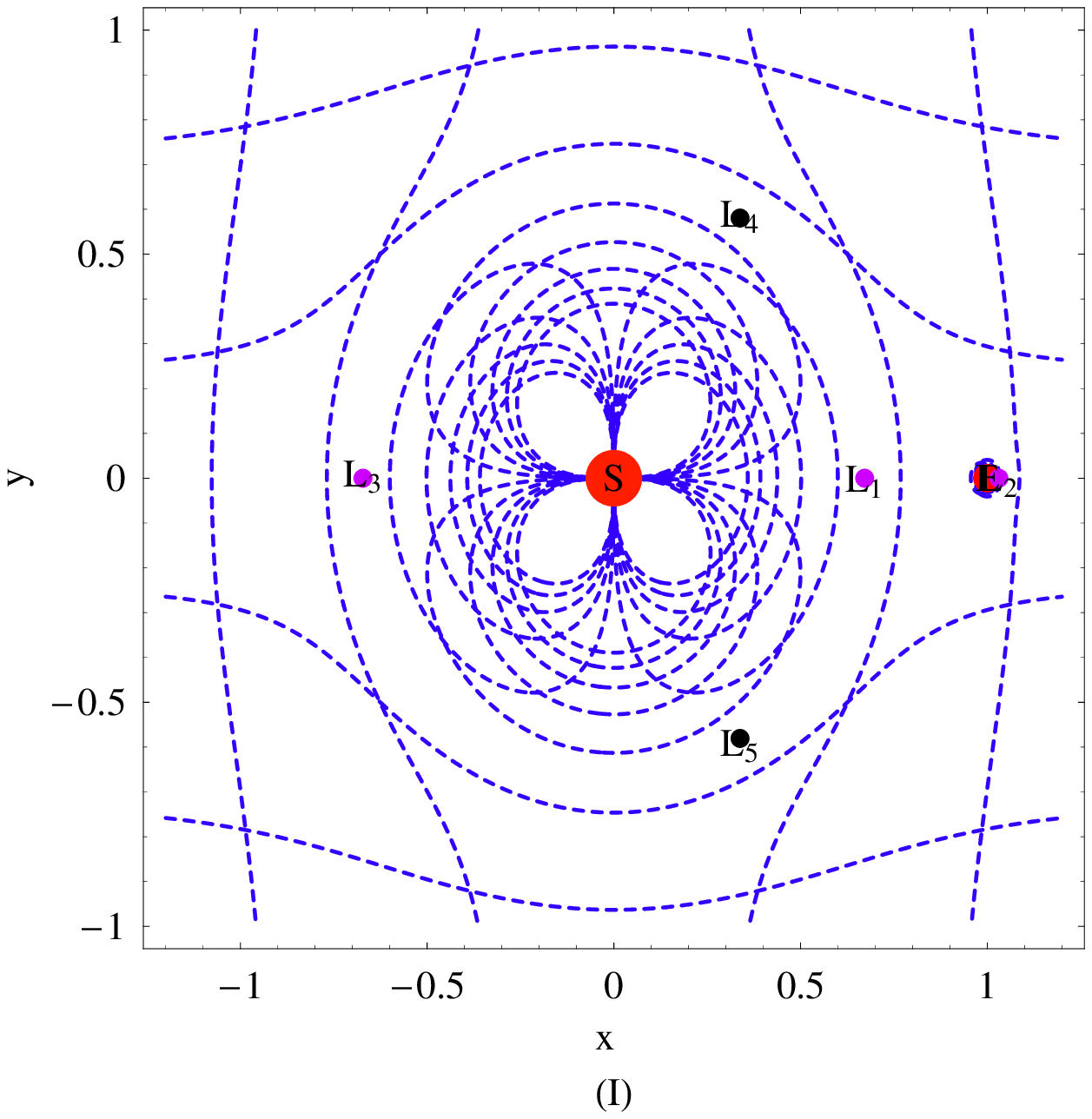}{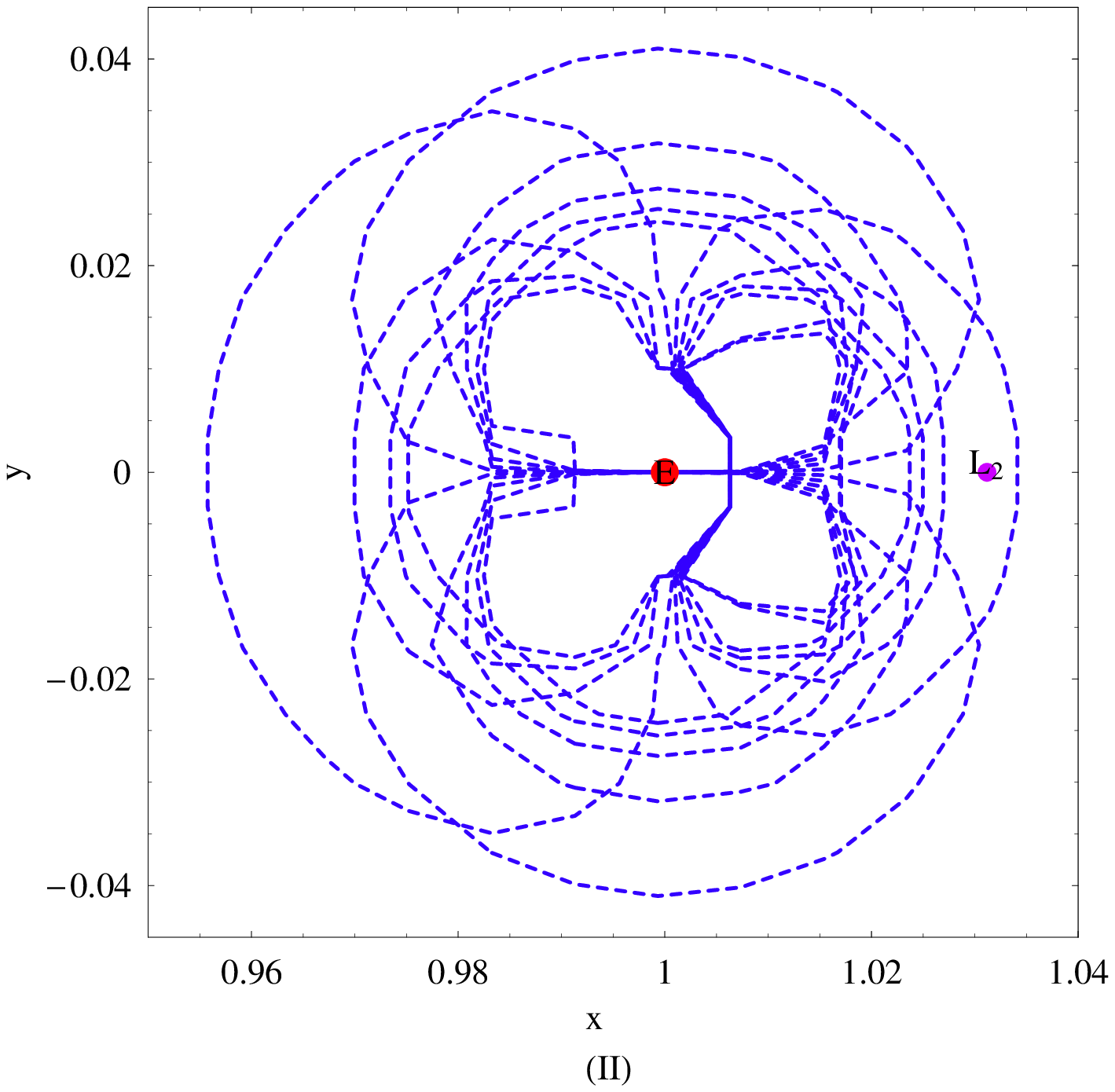}
   \caption{The position of equilibrium points  when T=0.1, $q_1=0.5$, $A_2=0.25$ and $M_b=0.25$, panel (I): Pink points  are collinear points and black  points are triangular points  (II): Position of $L_2$ with respect to  Earth is shown in zoom.}
   \label{fig:lpoints}
\end{figure}
The equilibrium points are shown in figure \ref{fig:lpoints} in
which two panels i.e. (I) pink points
correspond to the collinear points and black points correspond to the triangular points for the Sun-Earth system, whereas panel (II) show the zoom of the neighborhood of $L_2$. The numerical values of these points  are presented in Table
\ref{tab:lpts}. It is seen that the positions of  $L_1, L_3$ are shifted to rightward; $L_2, L_4$  are shifted to  leftward; and  $L_4$ is also shifted to downward  with respect to their positions in the classical problem. The nature
of the $L_5$ is not discussed in present model because it is  same as the nature of  $L_4$. But the detail behavior of the
$L_2$ with stability regions is discussed in sections  \ref{sec:TrjL2} \& \ref{sec:stbL2}.

\begin{table}[h]
\caption{Location of equilibrium points, when $T=0.1$,  $A_2=0.25$ and $M_b=0.25$. }
\begin{tabular}{|c l l| l l|}
\hline\hline
$q_1=0.75$& &  & $q_1=0.5$& \\
$L_i$ & x  & y  & x  & y \\
  \hline\hline
$L_1$& 0.844989 &0 &0.671768& 0  \\ $L_2$& 1.03519 &0 &1.03118&0  \\ $L_3$&-0.845423&0&-0.671796&0\\ $L_4$&0.419679&0.733898&0.337923& 0.580616 \\ \hline
\end{tabular}\label{tab:lpts}
\end{table}

\subsection{Comments on the Parameters}
\label{subsec:CmtPrmtr}
However,in general, it might be difficult to know the critical values of the parameters, but they could  be obtained  with the help of Interval Arithmetic(IA), which was introduced by \cite{mooreRE}. As per the IA, if $I_a=[a_1,a_2],I_b=[b_1,b_2]$  be  two intervals, then four basic arithmetic operations can be defined as:
\begin{itemize}
 \item Sum: $I_1+I_2=[a_1+b_1,a_2+b_2]$.
\item Difference: $I_1-I_2=[a_1-b_2,a_2-b_1]$.
\item Product:  $I_1\times I_2=[\mbox{min}(a_1b_1,a_1b_2,a_2b_1,a_2b_2),\mbox{max}(a_1b_1,a_1b_2,a_2b_1,a_2b_2) ]$.
\item Division: $I_1 / I_2=[\mbox{min}(a_1/b_1,a_1/b_2,a_2/b_1,a_2/b_2),\mbox{max}(a_1/b_1,a_1/b_2,a_2/b_1,a_2/b_2) ]$. The division by an interval containing zero is not defined in basic IA, so this case is  avoided.
\end{itemize}
It is supposed that  mass of the   Sun $m_1$ is  greater than  mass of the  Earth $m_2$, therefore $m_1\leq m_1+m_2$ and $m_2\leq m_1+m_2$. In other words,  $m_1$ lies in $ [\epsilon_1,m_1+m_2]$ and  $m_2$ lies in $[ \epsilon_2,m_1+m_2]$, where $\epsilon_1\geq  \epsilon_2\geq 0$. Using relation $\mu=\frac{m_{2}}{m_{1}+m_{2}}$, the domain of  mass parameter can be obtained as:
\begin{eqnarray}
 &&\mu \in \frac{[\epsilon_2,m_1+m_2]}{[\epsilon_1,m_1+m_2]+[\epsilon_2,m_1+m_2]},\\
\mbox{or} && \mu\in \left[\frac{\epsilon_2}{2},\frac{m_1+m_2}{2}\right].\end{eqnarray}
 From relation  $r_c^2=(1-\mu)q_1^{2/3}+\mu^2$,  the domain of mass reduction factor $q_1$ is given as:
\begin{eqnarray}
&&q_1^{2/3} \in \left[\frac{\epsilon_2^2-(m_1+m_2)^2}{1-\frac{(m_1+m_2)}{2}}, \frac{r^2-\epsilon_2^2}{1-\frac{(m_1+m_2)}{2}}\right].
\end{eqnarray}
And from relation $A_{2}=\frac{r^{2}_{e}-r^{2}_{p}}{5r^{2}}$, where  $r_e\in[\epsilon_4,r]$ and $r_e\in[\epsilon_5,r]$,   $\epsilon_4\geq \epsilon_5\geq 0$. The domain of oblateness coefficient can be obtained as:
\begin{eqnarray}
&&A_2 \in \frac{[\epsilon_4,r]^2-[\epsilon_5,r]^2}{5r^2}\\
\mbox{or}  && A_2\in\left[ \frac{\epsilon_4^2}{5r^2}-\frac{1}{5},\frac{1}{5}-\frac{\epsilon_5^2}{5r^2}\right]
\end{eqnarray}
Now from relation $n^{2}=1+\frac{3A_{2}}{2}+\frac{2M_b
r_c}{\left(r_c^2+T^2\right)^{3/2}}$, $n\in [0,2]$, $T=0.1$,  then the domain of  $M_b$  is obtained as $[0,\frac{33(r+0.01)^{3/2}}{20r}]$. In particular if $\epsilon_i=0(i=1,2,3,4,5)$,  $ m_1+m_2=1$ (unit of mass), $r_c=r=1$(unit of distance), then  it is obtained  $\mu\in[0,\frac{1}{2}]$,  $q_1\in[-2\sqrt{2},2\sqrt{2}]$, $A_2 \in\left[-\frac{1}{5},\frac{1}{5}\right]$, and $M_b\in[0,1.6748]$.
But in the present model  it is  considered that $0\leq F_p\leq F_g$,  $r_e\geq r_p$, so definitely $q_1$ lies in [0,1], $A_2\in[0,1/5]$ i.e $A_2=2.4337\times10^{-12}$.
\section{Trajectory of $L_2$}
\label{sec:TrjL2} 
\begin{figure}
    \includegraphics[scale=.7]{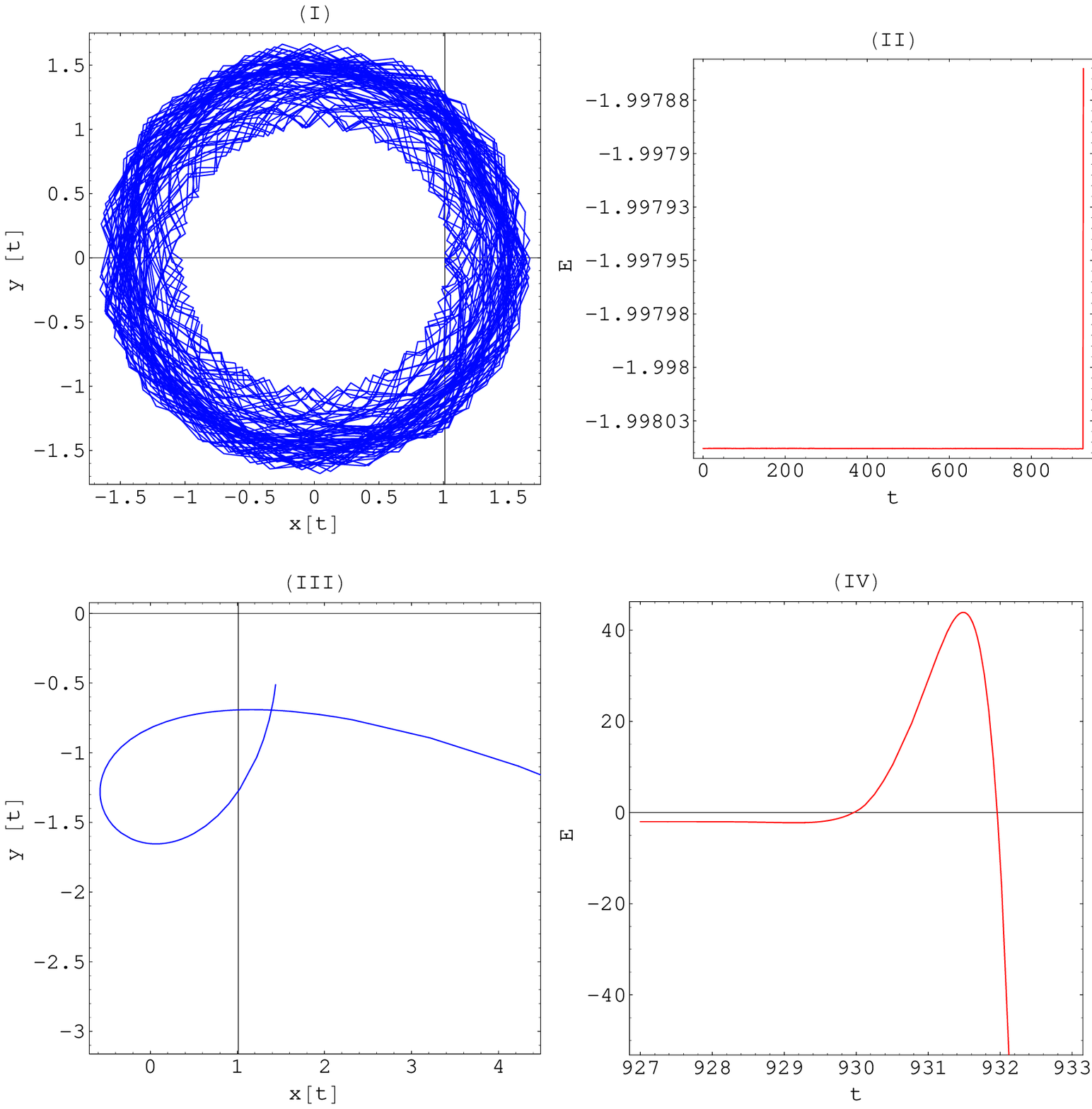}
   \caption{The Panels  (I-II):$0<t<927.3$ and (III-VI):$927.3<t<933$  in which (I and III) show the  trajectory of $L_2$, (II and IV) show  energy-versus time, when   T=0.1, $q_1=1$, $A_2=0$ and $M_b=0.25$.}
   \label{fig:trjq1a0m25}
\end{figure}

The equations (\ref{eq:Omegax}-\ref{eq:Omegay})
with initial conditions $x(0)=x(L_2), y(0)=0, x'(0)=y'(0)=0$ are
used to determine the trajectories of $L_2$ for different possible
cases. For plotting of the figures, the position of $L_2$ at $t=0$ is considered as the origin of coordinate axes. The figure
\ref{fig:trjq1a0m25}  show the trajectories of $L_2$ with four  panels when $q_1=1, M_b=0.25, A_2=0$. The panels (I-II): describe the case $0<t<927.3$. In panel(I) trajectory is  moving chaotically around  the  $L_2$ with $x \in
(-1.66639,1.66663)$, $y\in(-1.66885,1.6708)$ and in panel (II)  the 
energy integral  $E$ is oscillating  with negative values(approximate value is -1.99804). The panels (III-IV) are plotted for $927\leq t \leq 933$ for which  the trajectory departs from the point
 and energy integral becomes  negative for  $t<929.9$, after this time it becomes positive. The maximum value of  $E$ is 43.8974(at $t$=931.5). Also, when $t> 931.5$ then  $E$  is  found strictly decreasing. 
\begin{figure}
    \includegraphics[scale=.7]{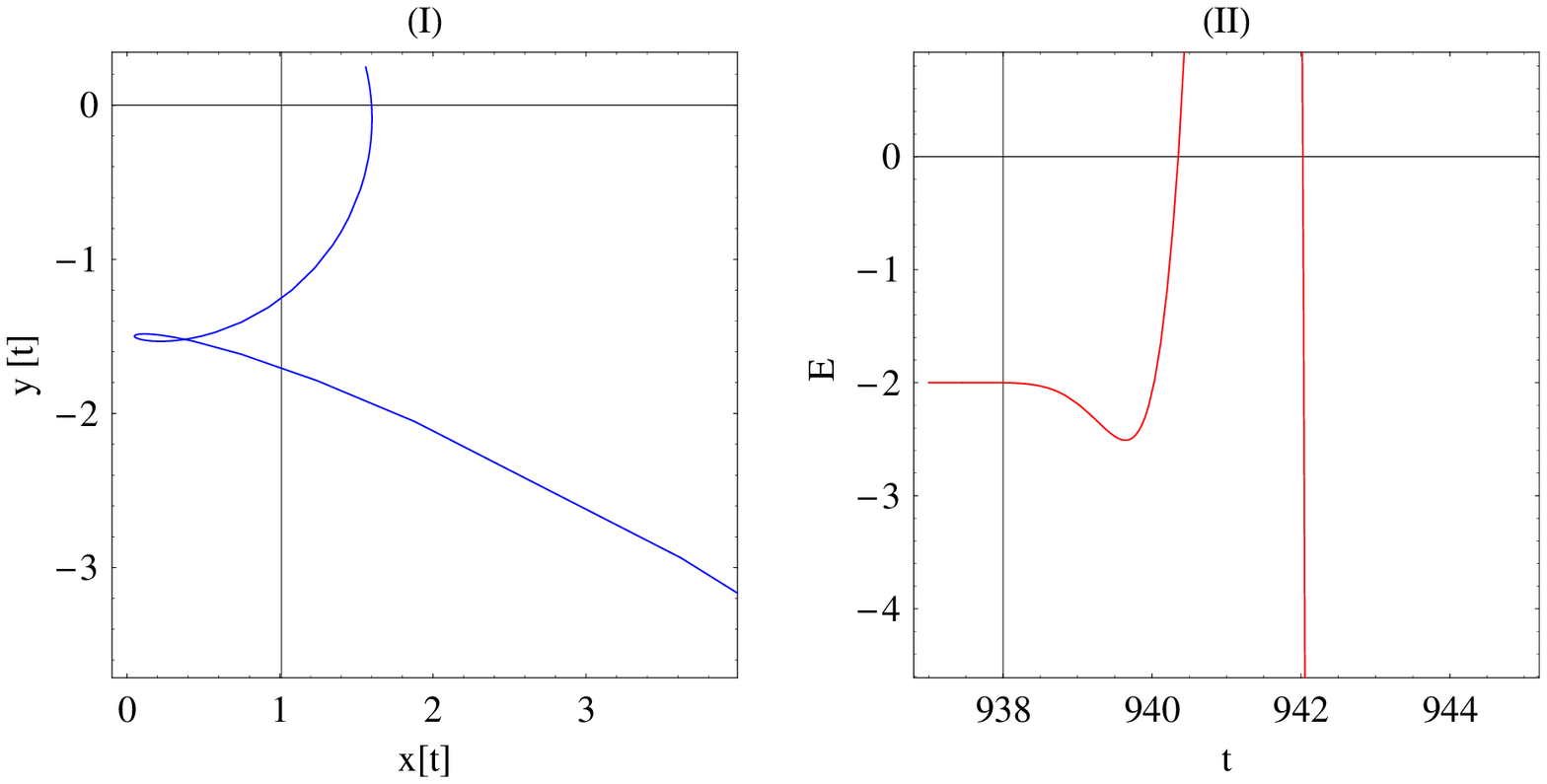}
   \caption{The Panels  (I-II):$937<t<945$ in which (I) show the  trajectory of $L_2$, (II) show  energy-versus time when  T=0.1, $q_1=1$, $A_2=0.0004$ and $M_b=0.25$.}
   \label{fig:trjq1a4m25}
\end{figure}
The effect of Earth oblateness  to the trajectory of $L_2$ is shown in figure \ref{fig:trjq1a4m25}  which is plotted for $q_1=1, M_b=0.25$ and $A_2=0.0004$.  The behaviour of trajectory is almost  same as in above case, and  integral of energy is  negative for $0<t<937.47$.
 It is clear from panels (I-II) the trajectory moves away from the Lagrangian point $L_2$  when 
$t\geq940$. Initially energy integral has negative values for time $0\leq t \leq 940$
 and it becomes positive for  time $940\leq t \leq 942.02$, which attains maxima  $E=25.2749$ at $t=941.6$, if $t\geq 942.03$ then $E<0$ is found.
\begin{deluxetable}{rrrrrrrrrr}
\tabletypesize{\scriptsize}
\rotate
\tablecaption{Maximum value of time $t_m$  for which  trajectory  of $L_2$ moves around the point when $A_2=10^{-j}$, pair ($q_1,M_b$).}
\tablewidth{0pt}
\tablehead{
\colhead{j} & \colhead{(1,0)} & \colhead{(1,0.25)} & \colhead{(1,0.50)} & \colhead{(0.75,0)} &
\colhead{(0.75,0.25)} & \colhead{(0.75,0.50)} & \colhead{(0.50,0)} &
\colhead{(0.50,0.25)} & \colhead{(0.50,0.50)}}
\startdata
  -10  &1331.93101& 922.79781& 788.32512 &180.74391& 503.43326 &505.77125&808.57010& 420.18206&409.85720\\
-9   &1293.30457& 922.27673&790.59409 &194.76252& 681.94769 &520.06460&420.18206&420.60738&401.85174 \\
-8  & 1298.97760& 921.81836&785.85800 &419.97449& 505.87022 &521.37550&407.38452& 407.38452&394.22863\\
-7  & 1245.34016& 927.64717& 787.31619&652.84935&656.81667&535.30463&414.11940 &414.11940&405.16373\\
-6  & 1270.43530& 929.71580&789.17206&606.81842& 600.2735&521.63088 &423.85070&423.85070&406.16332\\
-5  & 1.21501& 929.21181& 787.05192 & 388.61276 &638.00066&537.18908& 424.55070 &424.55069&399.15415\\
-4  & 0.43533& 924.27518& 89.10469 & 596.68529 &594.73322&516.01512&403.91467 &403.91466&399.15415\\
-3  & 0.13977& 0.18754& 790.81103&0.18391& 542.68447 &493.82489&394.12204&394.12204&422.23115\\
-2  & 0.04427&0.045214&0.04624& 0.04517&0.047163&0.04951& 0.05098 &0.05098&0.05088\\
-1   & 0.01400&0.01403&0.01406& 0.01403 &0.01408&0.01414& 0.01417 &0.01417&0.01417\\
\enddata
%% Text for table notes should follow after the \enddata but before
%% the \end{deluxetable}. Make sure there is at least one \tablenotemark
%% in the table for each \tablenotetext.
\label{tab:MaxTforL2}\end{deluxetable}
 The details of trajectory and energy are  presented in Table
\ref{tab:MaxTforL2}  for various values of parameters  and the  effect of parameters on the stability  is presented in Table \ref{tab:efctPara}, when   $A_2=2.4337\times10^{-12}$(for Sun-Earth system).    One can see
that  maximum value $t_m$ of time  for which trajectory moves around the points, which is an decreasing function of  $M_b$. When $A_2$ is very small in ($10^{-10}-10^{-3}$) the value of $t_m$  is initially decreasing function of $A_2$ and increasing function of $q_1$. It is obtained that the value of  $t_m$ is very small when $A_2>10^{-2}$.
\section{Stability of $L_2$}
\label{sec:stbL2} 
\begin{figure}
    \includegraphics[scale=.7]{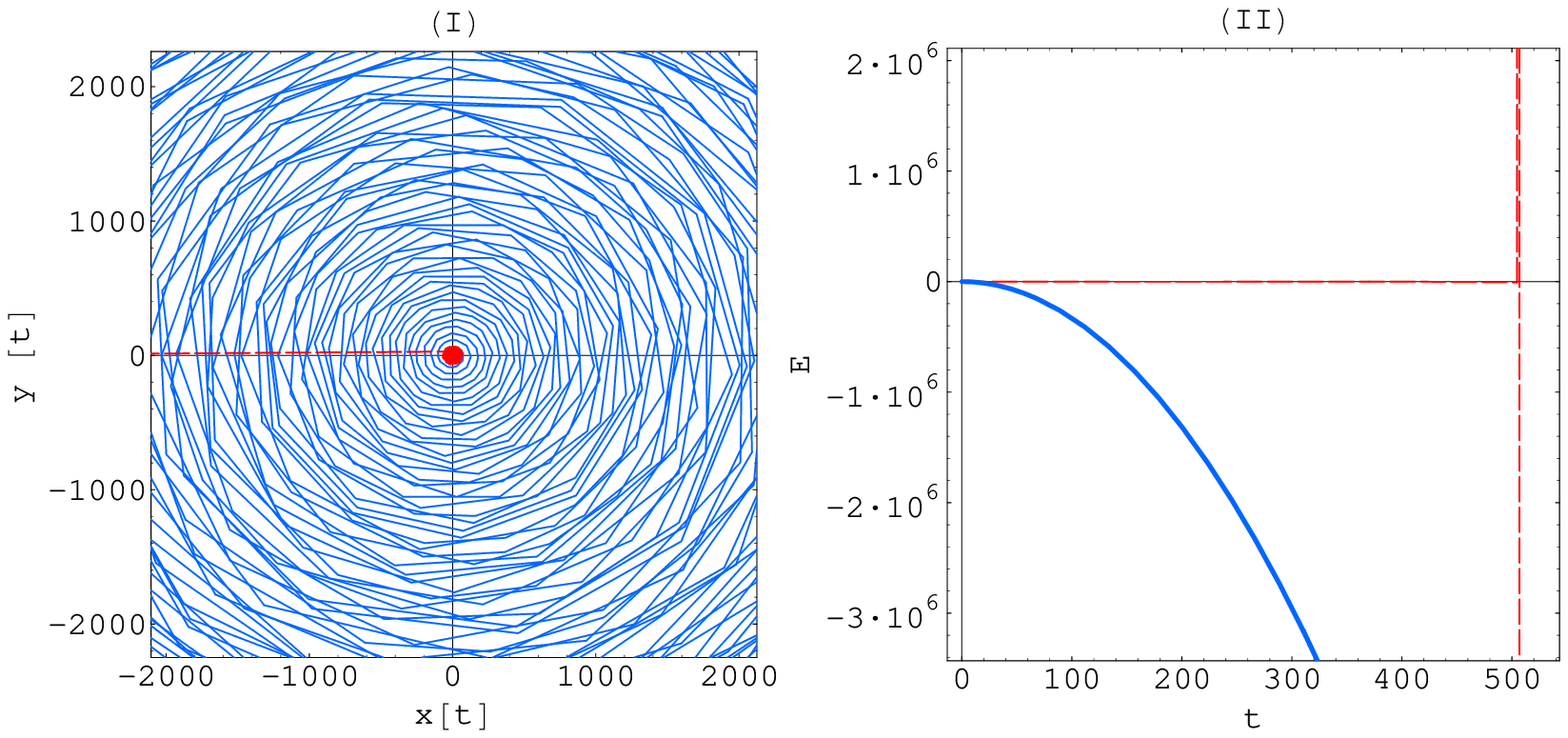}\\\includegraphics[scale=.7]{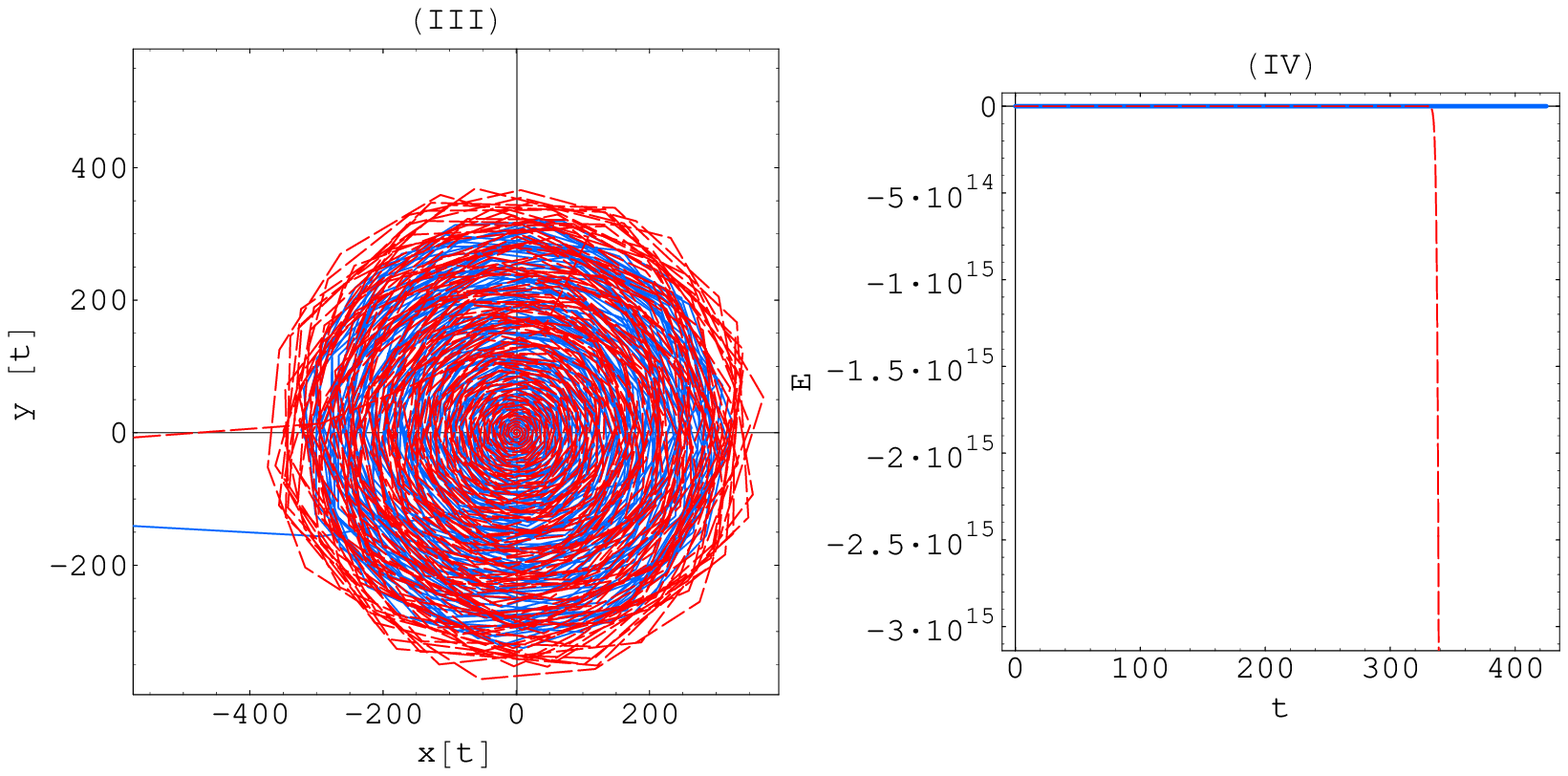}
   \caption{Panels (I \& III) show the stability regions of  $L_2$ (II \& IV )  show  the energy-versus time when  T=0.1, $q_1=0.75, 0.50$, $A_2=0.00$.}
   \label{fig:stbq75m2at}
\end{figure}

Suppose the coordinates $(x_1, y_1)$ of $L_2$ are
initially perturbed by changing \(x(0) = x_1+\epsilon \cos(\phi),
y(0) = y_1+\epsilon\sin(\phi)\), where \[ \phi
=\arctan\left(\frac{y(0)-y_1}{x(0)-x_1}\right)\in  (0, 2\pi),\] \( 0 \leq
\epsilon= \sqrt{(x(0) - x_1)^2 + (y(0) - y_1)^2} < 1\),  and $\phi$
indicates the direction of the initial position vector in the local
frame. If the $\epsilon = 0$ means there is no perturbation. It is
supposed that the $\epsilon = 0.001$ and the $\phi= \frac{\pi}{4}$
to examine the stability of $L_2$. Figure \ref{fig:stbq75m2at} show the
path of test particle and its energy with four panels i.e. the
panels (I\&III):$ q_1 = 0.75, 0.50,A_2 = 0.0$, in (I) trajectory of
perturbed $L_2$ moves in chaotic-circular path around initial
position without deviating far from it, then steadily move out of
the region. It is found that the test particle moves in the stability region and
returns repeatedly on its initial position. The blue solid curves
represent $M_b = 0.25$,  for panel(I) $t\leq 530$  and for panel (II) $t\leq 425$. The  red dashed  curves represent $M_b = 0.50$, for panel(III) $t\leq 510$  and panel (IV) $t\leq 350$.
\begin{figure}
    \includegraphics[scale=.7]{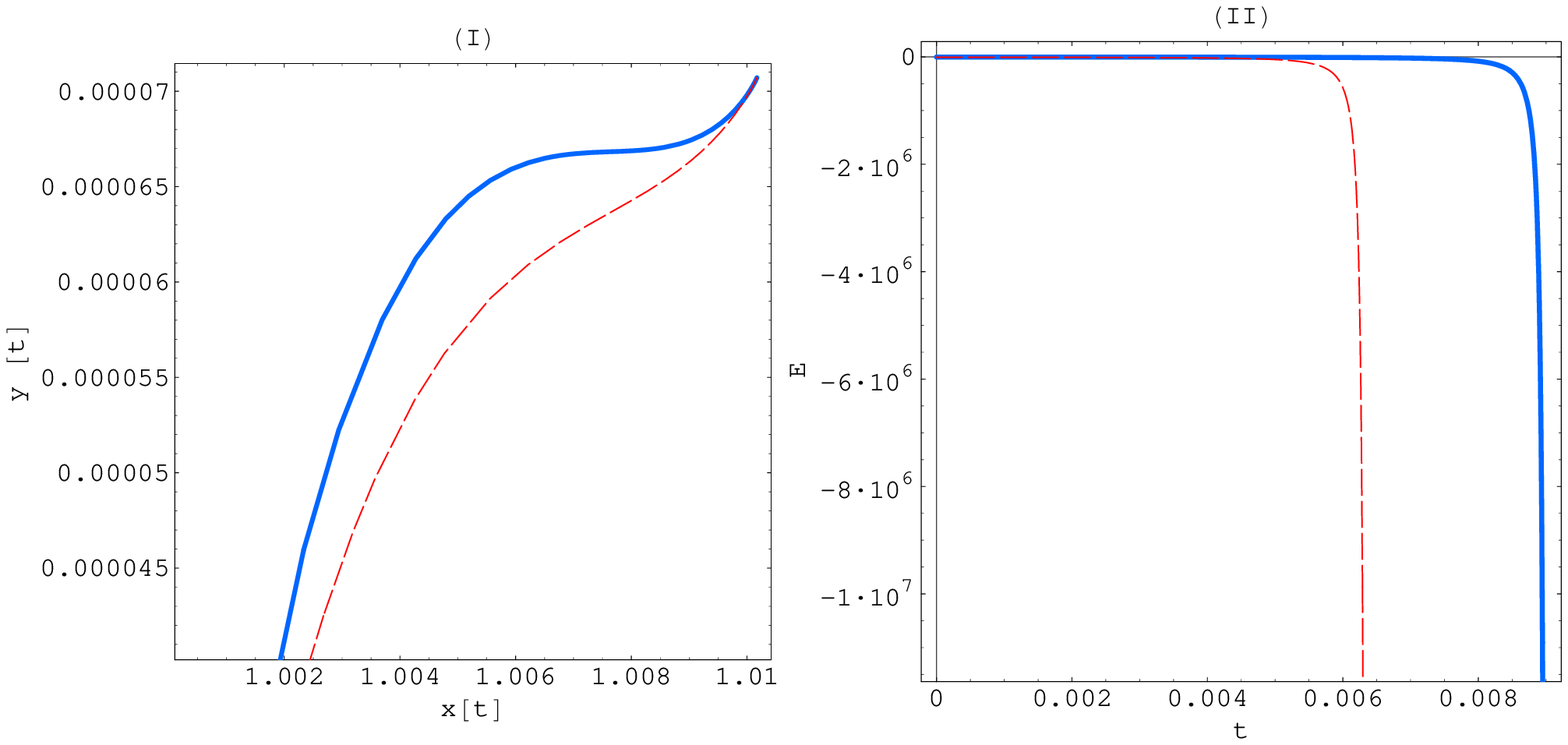}\\\includegraphics[scale=.7]{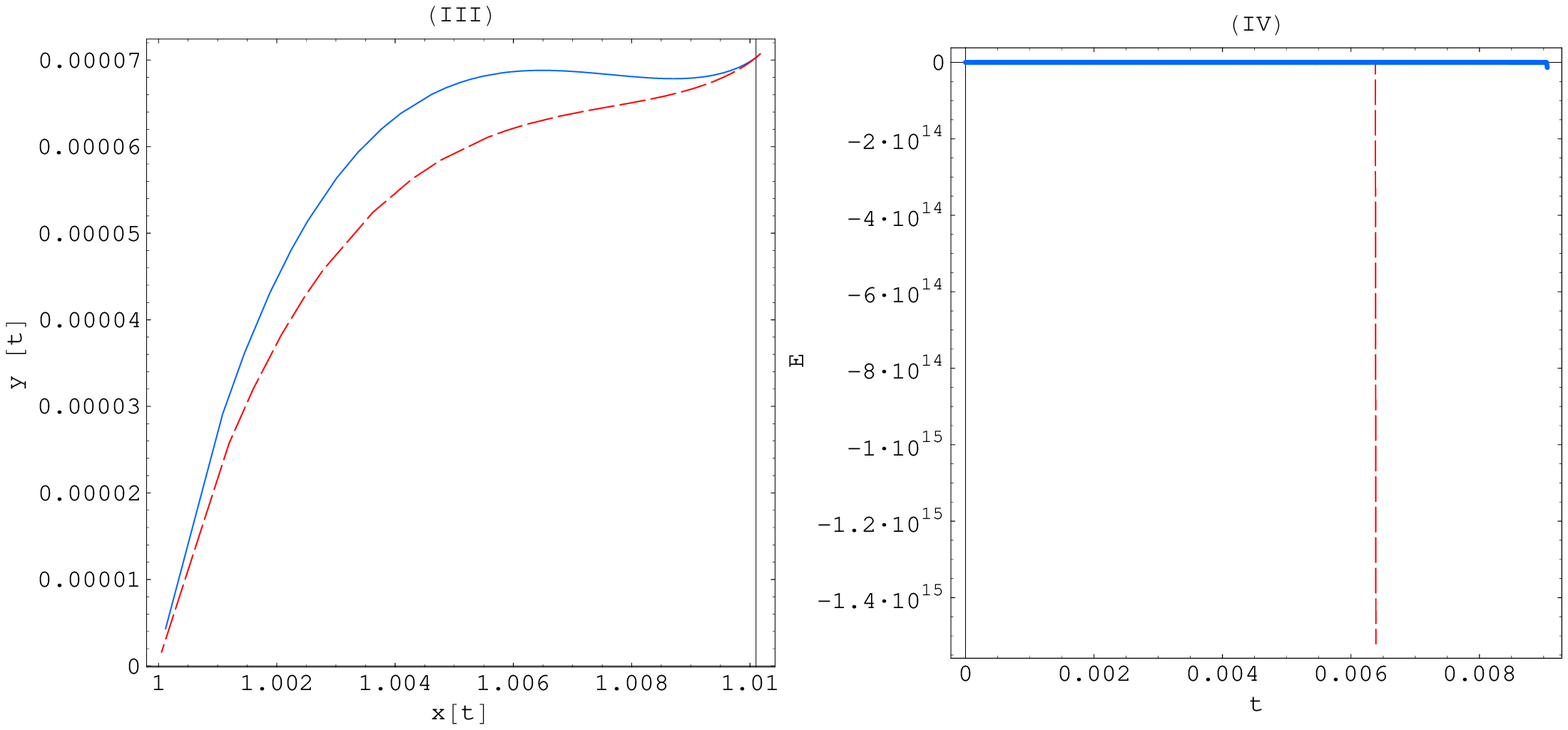}
   \caption{Show the  stability of $L_2$ with panels  (I-II):$q_1=0.75,M_b=.25$  and (III-IV):$q_1=0.50, M_b=0.25$ in which blue solid curves for $A_2=0.25$, red curves for $A_2=0.50$ when $T=0.1$.}
   \label{fig:stbOblt}
\end{figure}
The effect of oblateness of the second primary is shown in figure
\ref{fig:stbOblt} when $q_1=0.75 , M_b=0.25$. The panels (I\&III) show
the  trajectory of perturbed point $L_2$ and (II\&IV) show the energy
of that point. The blue solid lines correspond to $A_1=0.25$ and red dashed 
lines  correspond to $A_2=0.50$. One can see that the oblate effect is very
powerful on the trajectory and on the stability of $L_2$. When $A_2$ is very small 
the $L_2$ is asymptotically stable  for the value of $t$ which lies
within a certain interval. But if oblate effect of second primary is
grater than $10^{-2}$, the stability region of $L_2$ disappears.
\begin{center}
\begin{table}
\caption{Time $t_m$ for the stability of $L_2$ when   $T=0.1$,  $A_2=2.4337\times10^{-12}$}
\begin{tabular}{|l |l l l l l l|}
\hline\hline
$q_1$&$M_b$=0.0& $M_b$=0.20&$M_b=0.40$ &$M_b=0.60$ &$M_b=0.80$ &$M_b=1.00$ \\\hline\hline
1.00 &1330.94105&898.46095&740.63037&612.37135&476.86492&338.8331\\
0.90& 822.12754&846.66095&695.75819&475.72650&384.90335&315.45105\\ 
0.80&872.68679&680.99491&577.52196&428.50822&357.90646&320.62530\\ 
0.70&645.16838&563.21770&581.43694&388.86721&355.88230&313.83184\\
0.60&599.14033&534.73979&446.02727&379.57860&350.15274&326.96167\\
0.50&821.08169&559.32274&432.84188&381.03250&343.84316&320.76820\\
0.40&720.76190&95.04409&429.40509&380.36035&343.26450&317.89680\\
0.30&651.26546&491.31201&425.49945&397.40864&333.97100&307.68910\\
0.20&640.84770&485.92301&435.45741&379.56781&344.58453&317.24080\\
0.10&609.87381&505.53178&413.39252&379.19007&328.82013&313.27721\\
0.00&600.63232&478.59863&416.37905&390.05165&334.88598&308.42592\\
 \hline
\end{tabular}\label{tab:efctPara}
\end{table}
\end{center}
\section{Conclusion}
The numerical computation presented in the manuscript  provides
remarkable results to design trajectories  of Lagrangian point $L_2$
which helps us to make comments on the stability(asymptotically) of
the point. We obtained the intervals of the  time where trajectory
continuously  moves around the $L_2$,  does not deviate far from the
point but tend to approach (for some cases) it, the energy of
perturbed point is negative for these intervals, so we conclude that
the point is asymptotically stable. More over we have seen that after
the specific time intervals the trajectory of perturbed point departs
from the neighborhood and goes away from it, in this case the energy
also  becomes positive, so the Lagrangian point  $L_2$ is unstable.
Further the trajectories and the stability regions are affected by
the radiation pressure, the oblateness of the second primary and
mass of the belt.
\section*{Acknowledgments}
The author wishes to express his thanks to Indian School of Mines, Dhanbad (India), for providing financial support through Minor Research Project (No.2010/MRP/04/Acad. dated $30^{th}$ June 2010).  The author is also wishes to express his thanks to DST(Department of Science and Technology), Govt. of India for supporting this research through SERC-Fast Track Scheme for Young Scientist (DO.No.SR/FTP/PS-121/2009, dated $14^{th}$ May 2010).


\bibliographystyle{elsarticle-harv}

\begin{thebibliography}{16}
\expandafter\ifx\csname natexlab\endcsname\relax\def\natexlab#1{#1}\fi
\expandafter\ifx\csname url\endcsname\relax
  \def\url#1{\texttt{#1}}\fi
\expandafter\ifx\csname urlprefix\endcsname\relax\def\urlprefix{URL }\fi

\bibitem[{{Chermnykh}(1987)}]{Chermnykh1987}
{Chermnykh}, S.~V., 1987. {Stability of libration points in a gravitational
  field.} Vest. Leningrad Mat. Astron. 2, 73--77.

\bibitem[{{Farquhar}(1967)}]{Farquhar1967JSpRo}
{Farquhar}, R.~W., Oct. 1967. {Lunar Communications with Libration-Point
  Satellites}. Journal of Spacecraft and Rockets 4, 1383--1384.

\bibitem[{{Farquhar}(1969)}]{Farquhar1969AsAer}
{Farquhar}, R.~W., 1969. {Future Missions for Libration-Point Satellites}.
  Astronautics Aeronautics, 52--56.

\bibitem[{{Grebennikov} and {Kozak-Skoworodkin}(2007)}]{Grebennikov2007CMMPh}
{Grebennikov}, E.~A., {Kozak-Skoworodkin}, D., Sep. 2007. {Numerical estimates
  for stability domains of Lagrangian solutions to the restricted three-body
  problem}. Computational Mathematics and Mathematical Physics 47, 1477--1488.

\bibitem[{{Jiang} and {Yeh}(2004{\natexlab{a}})}]{Jiang2004IJBC}
{Jiang}, I., {Yeh}, L., Sep. 2004{\natexlab{a}}. {Dynamical Effects from
  Asteroid Belts for Planetary Systems}. International Journal of Bifurcation
  and Chaos 14, 3153--3166.

\bibitem[{{Jiang} and {Yeh}(2004{\natexlab{b}})}]{JiangYeh2004AJ}
{Jiang}, I.-G., {Yeh}, L.-C., Aug. 2004{\natexlab{b}}. {On the Chaotic Orbits
  of Disk-Star-Planet Systems}. \aj 128, 923--932.

\bibitem[{{Jiang} and {Yeh}(2006)}]{JiangYeh2006Ap&SSI}
{Jiang}, I.-G., {Yeh}, L.-C., Dec. 2006. {On the Chermnykh-Like Problems: I.
  the Mass Parameter {$\mu$} = 0.5}. \apss 305, 341--348.

\bibitem[{{Kushvah}(2008)}]{Kushvah2008Ap&SS318}
{Kushvah}, B.~S., Nov. 2008. {Linear stability of equilibrium points in the
  generalized photogravitational Chermnykh's problem}. \apss 318, 41--50.

\bibitem[{{Kushvah}(2009{\natexlab{a}})}]{Kushvah2009Ap&SS}
{Kushvah}, B.~S., Sep. 2009{\natexlab{a}}. {Linearization of the Hamiltonian in
  the generalized photogravitational Chermnykh's problem}. \apss 323, 57--63.

\bibitem[{{Kushvah}(2009{\natexlab{b}})}]{Kushvah2009RAA}
{Kushvah}, B.~S., Sep. 2009{\natexlab{b}}. {Poynting-Robertson effect on the
  linear stability of equilibrium points in the generalized photogravitational
  Chermnykh's problem}. Research in Astronomy and Astrophysics 9, 1049--1060.

\bibitem[{{Miyamoto} and {Nagai}(1975)}]{MiyamotoNagai1975PASJ}
{Miyamoto}, M., {Nagai}, R., 1975. {Three-dimensional models for the
  distribution of mass in galaxies}. \pasj 27, 533--543.

\bibitem[{{Moore}(1963)}]{mooreRE}
{Moore}, R.~E., 1963. Interval arithmetic and automatic error analysis in
  digital computing. Ph.D. thesis, Stanford, CA, USA.

\bibitem[{{Papadakis}(2004)}]{Papadakis2004A&A}
{Papadakis}, K.~E., Oct. 2004. {The 3D restricted three-body problem under
  angular velocity variation}. \aap 425, 1133--1142.

\bibitem[{{Papadakis}(2005)}]{Papadakis2005Ap&SS299}
{Papadakis}, K.~E., Sep. 2005. {Motion Around The Triangular Equilibrium Points
  Of The Restricted Three-Body Problem Under Angular Velocity Variation}. \apss
  299, 129--148.

\bibitem[{{Szebehely}(1967)}]{Szebehely1967}
{Szebehely}, V., 1967. {Theory of orbits. The restricted problem of three
  bodies}. New York: Academic Press.

\bibitem[{{Yeh} and {Jiang}(2006)}]{YehJiang2006Ap&SSII}
{Yeh}, L.-C., {Jiang}, I.-G., Dec. 2006. {On the Chermnykh-Like Problems: II.
  The Equilibrium Points}. \apss 306, 189--200.
\end{thebibliography}

\end{document}